\documentclass{emulateapj}
\usepackage{xspace}
\usepackage{soul}
\usepackage{color}

\usepackage{longtable}
\usepackage{threeparttable}

\newcommand{\Fermi}{\emph{Fermi}\xspace}

\newcommand{\Swift}{\emph{Swift}\xspace}

\shorttitle{Analysis of Sub-threshold Short Gamma-ray Bursts in Fermi GBM Data}

\begin{document}
\title{Analysis of Sub-threshold Short Gamma-ray Bursts in Fermi GBM Data}

\author{
D.~Kocevski\altaffilmark{1},
E.~Burns\altaffilmark{2},
A.~Goldstein\altaffilmark{3},
T. Dal Canton\altaffilmark{2},
M.~S.~Briggs\altaffilmark{4,6},
L.~Blackburn\altaffilmark{6},
P.~Veres\altaffilmark{5},
C.~M.~Hui\altaffilmark{1},
R.~Hamburg\altaffilmark{4},
O.~J.~Roberts\altaffilmark{3},
C.~A.~Wilson-Hodge\altaffilmark{1},
T.~Littenberg\altaffilmark{1},
V.~Connaughton\altaffilmark{3},
J.~Racusin\altaffilmark{7},
A.~von~Kienlin\altaffilmark{8},
E.~Bissaldi\altaffilmark{9}
}

\begin{abstract}
The \Fermi Gamma-ray Burst Monitor (GBM) is currently the most prolific detector of Gamma-Ray Bursts (GRBs). Recently the detection rate of short GRBs (SGRBs) has been dramatically increased through the use of ground-based searches that analyze GBM continuous time tagged event (CTTE) data. Here we examine the efficiency of a method developed to search CTTE data for sub-threshold transient events in temporal coincidence with LIGO/Virgo compact binary coalescence triggers. This targeted search operates by coherently combining data from all 14 GBM detectors by taking into account the complex spatial and energy dependent response of each detector. We use the method to examine a sample of SGRBs that were independently detected by the Burst Alert Telescope on board the \emph{Neil Gehrels} \Swift Observatory, but which were too intrinsically weak or viewed with unfavorable instrument geometry to initiate an on-board trigger of GBM.  We find that the search can successfully recover a majority of the BAT detected sample in the CTTE data. We show that the targeted search of CTTE data will be crucial in increasing the GBM sensitivity, and hence the gamma-ray horizon, to weak events such as GRB 170817A. We also examine the properties of the GBM signal possibly associated with the LIGO detection of GW150914 and show that it is consistent with the observed properties of other sub-threshold SGRBs in our sample. We find that the targeted search is capable of recovering true astrophysical signals as weak as the signal associated with GW150914 in the untriggered data.

\end{abstract}


\altaffiltext{1}{Astrophysics Office, ST12, NASA/Marshall Space Flight Center, Huntsville, AL 35812, USA}
\altaffiltext{2}{NASA Postdoctoral Program Fellow, Goddard Space Flight Center, Greenbelt, MD 20771, USA}
\altaffiltext{3}{Science and Technology Institute, Universities Space Research Association, Huntsville, AL 35805, USA}
\altaffiltext{4}{Space Science Department, University of Alabama in Huntsville, 320 Sparkman Drive, Huntsville, AL 35899, USA}
\altaffiltext{5}{Center for Space Plasma and Aeronomic Research, University of Alabama in Huntsville, 320 Sparkman Drive, Huntsville, AL 35899, USA}
\altaffiltext{6}{Harvard-Smithsonian Center for Astrophysics, 60 Garden St, Cambridge, MA 02138, USA}
\altaffiltext{7}{Goddard Space Flight Center, Greenbelt, MD 20771, USA}
\altaffiltext{8}{Max-Planck-Institut fu?r extraterrestrische Physik, Giessenbachstrasse 1, 85748 Garching, Germany}
\altaffiltext{9}{Istituto Nazionale di Fisica Nucleare, Sezione di Bari, I-70126 Bari, Italy}

\section{Introduction} \label{sec:intro} 

The detection of gravitational waves (GW) from compact binary mergers by LIGO and Virgo has ushered in a new era in time-domain and multi-messenger Astronomy.  The detection of GW170817 in gravitational waves (GWs) by LIGO/Virgo \citep{GW170817} and GRB~170817A in gamma rays by the Fermi Gamma-ray Burst Monitor (GBM) \citep{GBM_only_paper} and INTEGRAL SPI-ACS \citep{Savchenko2017} resulted in follow-up observations across the electromagnetic spectrum \citep{GW170817-MMAD}. Among the many important discoveries enabled from this single event, the detection of coincident gamma-ray emission provided the first direct evidence for the long suspected connection between short gamma-ray bursts (SGRBs) and neutron star binary coalescence events \citep{jointpaper}.

The complementary information encoded in the electromagnetic (EM) signal associated with GW170817 showed that such observations can provide essential astrophysical context to GW detections. The GBM observations constrained the prompt energetics of the associated SGRB (GRB~170817A), revealed a thermal component in the prompt gamma-ray spectrum, and allowed for the tightest constraints on the speed of gravity \citep{GBM_only_paper, jointpaper}. Such joint LIGO/Virgo-GBM detections also provide immediate confirmation of the GW candidate and can reduce the region for follow-up by combining the independent localizations. Joint searches of GW strain data and GBM data can also provide increased confidence in weak GW detections, increasing the effective detection distance in both instruments.

The GBM is currently the most prolific detector of gamma-ray bursts (GRBs), making it the premiere instrument with which to search for coincident EM emission from GW detections. Consisting of an array of scintillation detectors observing the entire unocculted sky, GBM autonomously triggers on board to $\sim$240 GRBs per year, $\sim$40 of which are SGRBs. The instrument localizes these bursts to an accuracy of a few degrees, providing spectral information and high temporal resolution \citep{Meegan2009}, with which to perform detailed spectral modeling of short transient events in the 8 keV to 40 MeV energy range.

Recently the GBM detection rate of SGRBs has been increased dramatically through the use of ground-based searches to analyze GBM continuous time tagged event (CTTE) data. These offline searches employ sophisticated analysis methods that are not achievable in real time due to the limited computational resources available on the spacecraft. This makes these searches far more sensitive to SGRBs that are too intrinsically weak or distant to trigger GBM on board, or have a poor viewing geometry such that no, or only one, detector is brightly illuminated, thus failing the on-board triggering requirement of detections in at least 2 detectors. An offline blind search of the CTTE data, which uses a multi-detector rate analysis similar to that employed by the GBM flight software, but over a much larger range of energy bands and timescales, has been shown to find an additional $\sim$80 SGRB candidates per year. This technique was pioneered for the sub-threshold search of Terrestial Gamma-ray Flashes (TGFs) \citep{Briggs2013}, which are much shorter than SGRBs, resulting in substantial detection improvements compared to the in-orbit method, increasing from $\sim30$ to $\sim800$ TGFs per year~\citep{Roberts2018}.

Although the blind search examines a much larger parameter space than the analysis performed by the flight software, the search inherently still treats each GBM detector separately. In order to capitalize on the increased sensitivity that would be obtained through a coherent search of multiple-detector data, the GBM team developed a method to compare model predictions from a putative source on the sky to the observed signal in each detector~\citep{Blackburn2015, Goldstein2016}. The search uses the directionally dependent response of each detector to estimate the expected count rate from such a source on a grid of possible sky locations before marginalizing over either a uniform sky prior or a seeded localization probability map.  The expected counts as a function of energy are then compared to the observed counts, taking into account a modeled background component.  By combining the likelihood obtained from each detector comparison, the method allows for a much deeper search of the GBM data compared to treating each detector separately.

This targeted search was used to look for a candidate counterpart to the first direct observation of a binary black hole coalescence event, GW150914 (Abbott et al., 2016). That search, utilizing a coherent analysis over all 14 GBM detectors, resulted in a low-significance, spectrally hard candidate starting $\sim0.4$ s after the LIGO trigger time that is observationally consistent with a weak short GRB arriving at a poor geometry relative to the GBM detectors \citep{Connaughton2016}. The nature of this source, referred to here as GW150914-GBM, has resulted in a vigorous debate within the gamma-ray community (see \citealt{savchenko2016integral} and \citealt{greiner2016fermi} for the initiation of the controversy, and \citealt{2018arXiv180102305C} for a detailed response). Regardless of its true nature, GW150914-GBM highlights the ability of the targeted search to detect interesting sub-threshold events hidden in the GBM data that warrant further study.

We use the GBM offline targeted search to examine a sample of SGRBs that were independently detected by the Burst Alert Telescope (BAT) on board the \emph{Neil Gehrels} \Swift Observatory, a subset of which did not initiate an on-board trigger of GBM. We use this sample to examine the efficiency with which the targeted search can recover true astrophysical signals in the GBM CTTE data.  We compare the properties of these sub-threshold signals to the properties of SGRBs that triggered both BAT and GBM, as well as GRB 170817A and GW150914-GBM. 
 
We describe the BAT and GBM instruments in $\S$2, the sample selection in $\S$3, the data analysis method in $\S$4, the results of the analysis in $\S$5.  We discuss the results of the analysis in $\S$6, and conclude in $\S$7.

\section{Instrument Overviews}
\subsection{\emph{Neil Gehrels} \Swift Observatory}

The \emph{Neil Gehrels} \Swift Observatory consists of the BAT \citep{Barthelmy05}, the X-Ray Telescope (XRT) \citep{Burrows05}, and the UltraViolet Optical Telescope (UVOT) \citep{Roming05}. The BAT is a wide field, coded-mask gamma-ray telescope, covering a FoV of 1.4 sr and an imaging energy range of 15--150 keV. The instrument's coded-mask allows for positional accuracy of 1--4 arcminutes within seconds of a burst trigger. The XRT is a focusing X-ray telescope covering an energy range from 0.3--10 keV and providing a typical localization accuracy of $\sim$1--3 arcseconds.  The UVOT is a Ritchey-Chretien telescope that provides optical and ultraviolet photometry and grism spectroscopy with sub-arcsecond positional accuracy of the long-lived afterglow counterparts to the prompt emission from GRBs.  

\Swift operates autonomously in response to BAT triggers of new GRBs, automatically slewing to point the XRT and UVOT at a new source within 1--2 minutes.  Data are promptly downlinked to the ground, and localizations are available from the narrow-field instruments within minutes (if detected).  \Swift continues to follow-up GRBs as they are viewable outside of observing constraints and the observatory is not in the South Atlantic Anomaly (SAA) for at least several hours after each burst, sometimes continuing for days, weeks, or even months if the burst is bright and of particular interest for follow-up.

\subsection{\Fermi Gamma-ray Burst Monitor}

The GBM on board the \Fermi Gamma-Ray Space Telescope is composed of fourteen scintillation detectors designed to study the gamma-ray sky in the $\sim8$ keV to 40 MeV energy range \citep{Meegan2009}. Twelve of the detectors are semi-directional sodium iodide (NaI) detectors, which cover an energy range of 8--1000 keV, and are configured to view the entire sky unocculted by the Earth. The other two detectors are composed of bismuth germanate (BGO) crystals, covering a higher energy range of 200 keV to 40 MeV, and are placed on opposite sides of the spacecraft. Incident gamma-rays interact with the NaI and BGO crystals creating scintillation photons, which are collected by attached photomultiplier tubes and converted into electronic signals. The signal amplitudes in the NaI detectors have an approximately cosine response relative to the angle of incidence $\theta$, and relative rates between the various detectors are used to reconstruct source locations.  

The GBM flight software continually monitors the detector rates and triggers when a statistically significant rate increase occurs in two or more NaI detectors. Currently, 28 combinations of timescales and energy ranges are tested; with the first combination tested by the flight software that exceeds the predefined threshold (generally 4.5$\sigma$) being considered the trigger. 

Several data types are continuously produced by the GBM flight software by binning the observed counts into predefined timescales. These include CTIME (0.256 s over 8 energy channels) and CSPEC (4.096 s over 128 energy channels). The resolution of both these data types increase to 0.064 s and 1.024 s following an on-board trigger, respectively. During the first few years of the \Fermi mission, unbinned time-tagged event (TTE) data collected over 128 energy channels, were only produced around on-board triggers. A flight software update in 2012 November enabled the collection and downlinking of continuous TTE (CTTE) data for offline analysis. The CTTE data type is especially useful as it provides arrival time information for individual photons at 2 $\mu$s precision over 128 energy channels and is the basis for the offline sub-threshold analyses developed by the GBM team. 

\section{Sample Definition} \label{sec:SampleDefinition}

We compiled a sample of all SGRBs detected by BAT and observed by GBM between the beginning of \Fermi~science operations on 2008 August 4 and 2017 August 4.  We define a GRB being observed by GBM as occurring in the region of the sky within 113$^\circ$ of the Earth's zenith at the time of the BAT detection and at a time when \Fermi was not in the SAA. The resulting sample includes a total of 44 BAT detected SGRBs observed by GBM. Of these bursts, 33 also triggered GBM.  The remaining 11 bursts were detected by the BAT, but did not result in an on-board trigger of the GBM.  We define these 11 bursts as the GBM sub-threshold population.

For the 44 bursts in this sample, we utilized the Third \Swift BAT GRB Catalog \citep{Lien2016} to extract the relevant temporal and spectral properties of each burst as inferred from the BAT observations.  These include the burst duration ($T_{90}$), peak photon flux, energy fluence, and the best fit spectral model.  Since the peak flux and fluence estimates depend on the assumed spectral model, we selected the values associated with the best fit spectral model for each burst.

\section{Analysis}

For each \Swift detected burst in our sample, we utilized the GBM targeted search to examine a $\pm$ 5 s window of GBM data centered on the BAT trigger time (T$_0$) to identify coincident signals in GBM.  The GBM targeted search is described in greater detail in \citet{Blackburn2015} and \citet{Goldstein2016}, but is summarized here for convenience.  For SGRBs prior to 2013 the search was performed on 0.256 s resolution CTIME data from all GBM detectors (NaI and BGO) covering nested timescales $T$ between 0.256 s and 8.192 s in duration.  The analysis was performed over 8 energy channels with up to 4 phase steps (limited by the CTIME temporal resolution), so as to ensure that bin boundaries do not mask the existence of a possible signal.  For SGRBs that occurred after 2013 the search was performed on CTTE data from all GBM detectors (NaI and BGO), covering nested $T$ ranging between 0.064 and 8.192 s in duration.  The analysis of the CTTE data was performed over 8 energy channels for 16 phase steps per timescale, limited to a minimum phase step of 0.064 s.

For the bursts that occurred prior to 2013, the targeted search estimates the background of each detector by fitting a polynomial to data from $-10T$ to $+10T$, excluding the interval from $-3T/2$ to $+5T/2$.  This background estimation is independently determined for each channel in each detector. For bursts after 2013, the background is estimated using a more sophisticated unbinned maximum Poisson estimation described in \citet{Goldstein2016}.  

The search uses the directionally-dependent response of each detector to estimate the expected count rate due to a putative source on the sky at a target time. The presence of such a source can then be tested by comparing the expected count rate to the observed counts.  In order to estimate the expected counts spectrum from the source, the search relies on three template photon spectra, which are folded through each of the GBM detector responses.  These templates, designated `soft', `normal', and `hard', consist of two Band functions and a power-law with an exponential high-energy cutoff, respectively, and are intended to represent a range of GRB spectra observed by GBM.  Since the location of the source is not assumed a priori, the expected counts are estimated over a 1$^\circ$ grid of possible locations on the sky. The expected counts per energy channel for each detector and spectral template combination, and for each location in the sky, is then compared to the observed counts, taking into account the modeled background.  The probability of measuring the observed counts $d$, from a source of amplitude $s > 0$, in the presence of the estimated background $n$ is given by:

\begin{equation}
P(d|H_1) = \prod_i \frac{1}{\sqrt{2\pi}\sigma_{d_{i}}} \exp{\bigg(-\frac{(\widetilde{d_i}-r_i s)^2}{2\sigma^2_{d_i}}\bigg)}
\end{equation}

\noindent where the product is carried out over every detector-time-energy combination and $\widetilde{d_i} = d_{i} - \langle n_i \rangle$ represents the background-subtracted data, $\sigma_{d_{i}}$ represents the standard deviation of the expected data (background+signal), $r_i$ represents the location and spectrally dependent instrument response, and $s$ is the true source amplitude in the observer frame. This can be compared to the probability that the observed counts are simply due to background ($s=0$), given by:

\begin{equation}
P(d|H_0) = \prod_i \frac{1}{\sqrt{2\pi}\sigma_{n_{i}}} \exp{\bigg(-\frac{\widetilde{d_i}^2}{2\sigma^2_{n_i}}\bigg)}
\end{equation}

\noindent where $\sigma_{n_{i}}$ represent the standard deviation of the background. 

A likelihood ratio is employed to compare the presence of a signal $H_1$ to the null hypothesis $H_0$ of pure background resulting in a test statistic with which to gauge the significance of a putative source:

\begin{equation}
\mathcal{L} = \ln{\frac{P(d|H_1)}{P(d|H_0)}} = \sum\limits_{i} \bigg[ \ln{\frac{\sigma_{n_{i}}}{\sigma_{d_{i}}}} + \frac{\widetilde{d_i}^2}{2\sigma_{n_{i}}^{2}} -\frac{(\widetilde{d_i}-r_i s)^2}{2\sigma^2_{d_i}}\bigg]
\label{Eq:logLikelihoodRatio}
\end{equation}

\noindent 
For a single location on the sky, we can maximize the log likelihood ratio by varying the source amplitude $s$. This entails minimizing the magnitude of the third term in equation \ref{Eq:logLikelihoodRatio}, at which point the test statistic is proportional to the signal to noise ratio (SNR) of the source.  Because the spectrum and sky location of the source are unknown a priori, we marginalize over these parameters to produce a final amplitude-marginalized log likelihood ratio $\mathcal{L}$.  According to Wilks' theorem, $\mathcal{L}$ should be distributed approximately as $\chi^{2}$, so we choose to reject the null hypothesis when $\mathcal{L} > 9$, roughly equivalent to a $3 \sigma$ rejection criteria for a single degree of freedom. Any detected fluctuations within the search window, characterized by their time and duration, are then ranked by their likelihood ratios.  The source with the highest likelihood ratio is chosen as the most significant detection for each burst in our sample. 

The likelihood ratios over the 1$^\circ$ grid of possible locations for the timescale and bin phasing that maximized the source significance provides a posterior probability distribution over the sky with which to localize the GBM signal.  From this probability distribution, we estimate the statistical 90$\%$ credible localization regions for each source. We convolve this region with a 7.6$^\circ$ Gaussian systematic uncertainty, determined from a comparison of SGRBs that triggered GBM and which have accurate localizations determined from other instruments. This provides a 90$\%$ credible localization regions that incorporates both statistical and systematic uncertainties.

Finally, testing for the presence of a signal above background over such a wide range of timescales, bin phases, and sky locations introduces a non-negligible number of search trials. These trials increase the probability that the search could find a statistically significant detection by chance alone due to the size of the parameter space being examined.  Unfortunately, the data in each of our trials are not statistically independent, as the various timescales and bin phases are nested and the sky locations overlap due to the large FOV of the detectors. The GBM data also exhibits considerable non-Gaussian backgrounds, including contributions from both non-Gaussian noise and real astrophysical events. These factors preclude a simple analytic estimation of a stricter significance threshold for individual comparisons, so as to compensate for the number of inferences being made. Instead, we form a false alarm rate (FAR) distribution to quantify the frequency of occurrence of short transients in the GBM data.  By its nature, the FAR distribution includes transient signals that are due to both statistical fluctuations and background astrophysical sources.  The FAR distribution that we employ for this purpose is formed by applying the targeted search over $\sim$10$^{5}$ s of GBM data. A post-trials chance association, or false-alarm probability (FAP), can then be estimated by calculating the Poisson probability of having a signal of rate $\lambda_{c}$ occur by chance within a given time window $P(\Delta~t<T) = 1 - e^{-3\lambda_{c}T}$, where the extra factor of three accounts for trial factors introduced by the three separate spectral templates employed by the search. This formulation of the FAP assumed a uniform probability across the search window and does not take into account the proximity of the detected signal to the seeded search time. 
\section{Results}

The application of the GBM targeted search on a $\pm$ 5 s window centered at the BAT trigger time resulted in the detection of a candidate source with $\mathcal{L} > 9$ in 42 of the 44 bursts in our sample.  The two remaining bursts, GRBs 090815C and 150728A, fall short of this detection threshold, with $\mathcal{L}$ values of 7.13 and 6.30 respectively.  Of the three spectral templates used by the targeted search, the medium template resulted in the highest fraction (45$\%$) of detections in the sample.  The remaining two templates, soft and hard, represented the roughly 20$\%$ and 34$\%$ of the sample, respectively.

\begin{figure}
\centering
\includegraphics[width=0.5\textwidth]{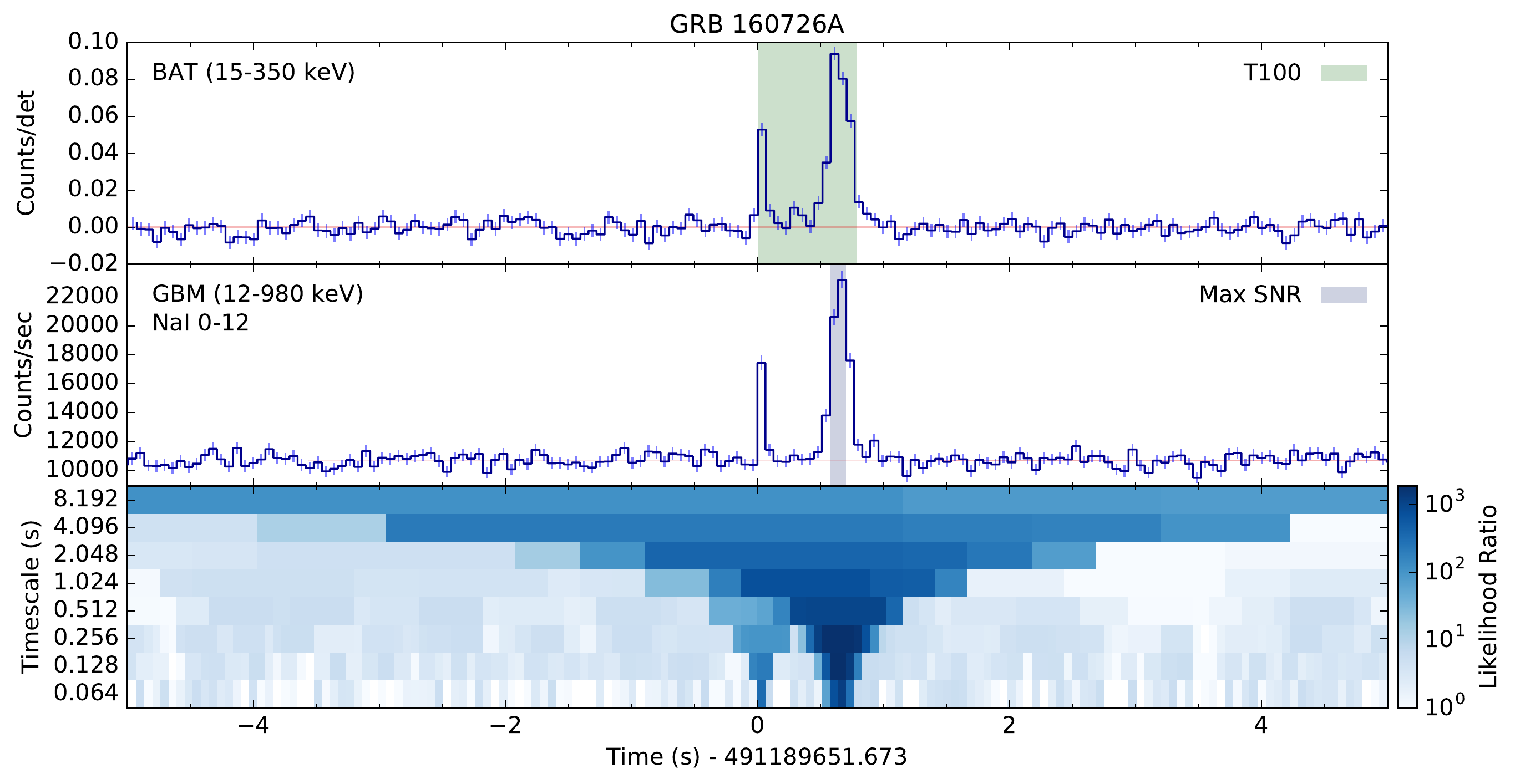}
\includegraphics[width=0.5\textwidth]{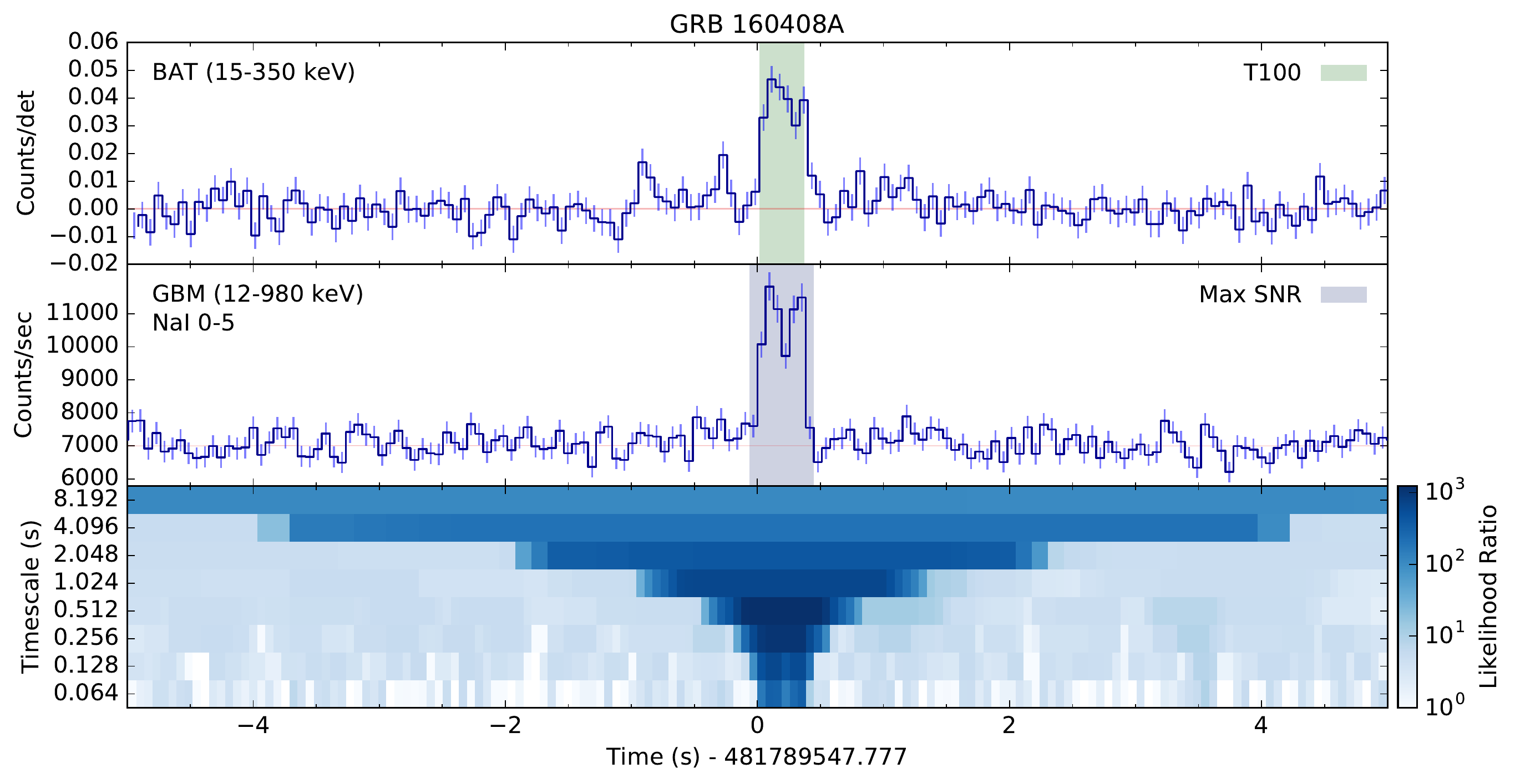}
\caption{Example BAT (top panels) and GBM (middle panels) light curves for 2 GRBs that triggered both instruments. The event display (lower panels) show the likelihood ratios, or the significance of the signal above the local background, for a range of bin timescales and phases.  The timescale that maximizes the signal significance in the GBM data is shown by the blue shaded region in the middle panels. Likewise, the phase of the GBM light curve is set to the value that maximize the log likelihood ratio.}
\label{Fig:LightCurves_TriggeredExamples}
\end{figure}

Example BAT and GBM light curves for two bursts, GRBs 160726A and 160408A, which triggered both GBM and BAT are shown in Figure \ref{Fig:LightCurves_TriggeredExamples}.  Each figure also contains an event display panel summarizing the likelihood ratios obtained for the range of timescales and bin phases examined by the targeted search. GRBs 160726A and 160408A are well detected above background in both instruments, and this is reflected in their event display panels, which show elevated likelihood ratios due to counts in excess of the estimated background for a particular time bin.  Both bursts were sufficiently bright to contribute excess counts on all timescales analyzed by the search, resulting in the cascading pattern of elevated likelihood values down to the smallest timescales surrounding T$_0$. The timescale that provided the highest likelihood ratio, and hence the greatest SNR in the multi-detector data, is highlighted as the blue shaded region in the middle panel of each subplot.  This detection window can be compared to the Swift BAT T$_{100}$ duration for each burst, shown as the green shaded region in the top panel. In the case of GRB 160726, the detection window is offset from the BAT trigger time by $>1$ s because of significant substructure in the burst light curve, where BAT triggered on a precursor to the primary emission episode of the burst. 

\begin{figure}
\centering
\includegraphics[width=0.5\textwidth]{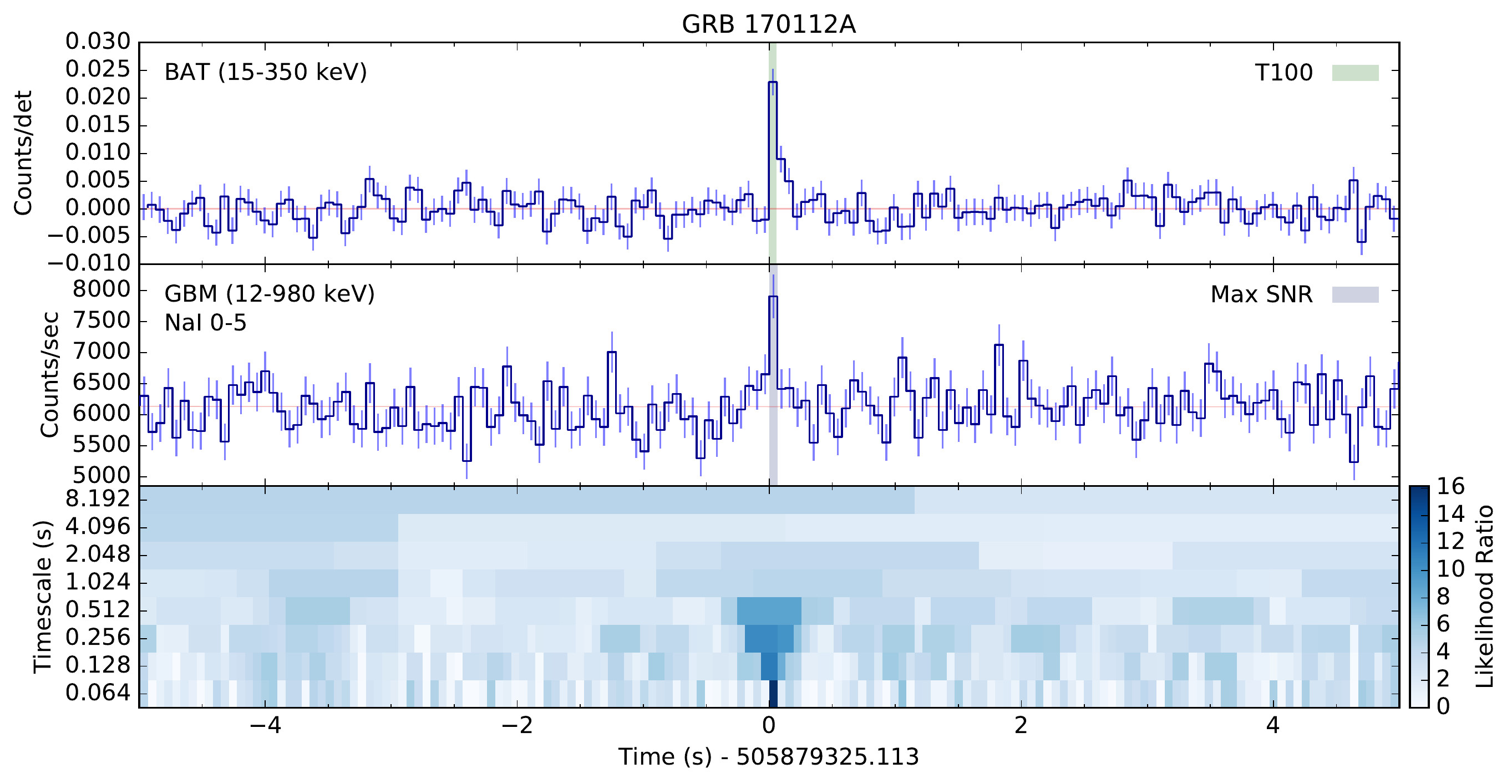}
\includegraphics[width=0.5\textwidth]{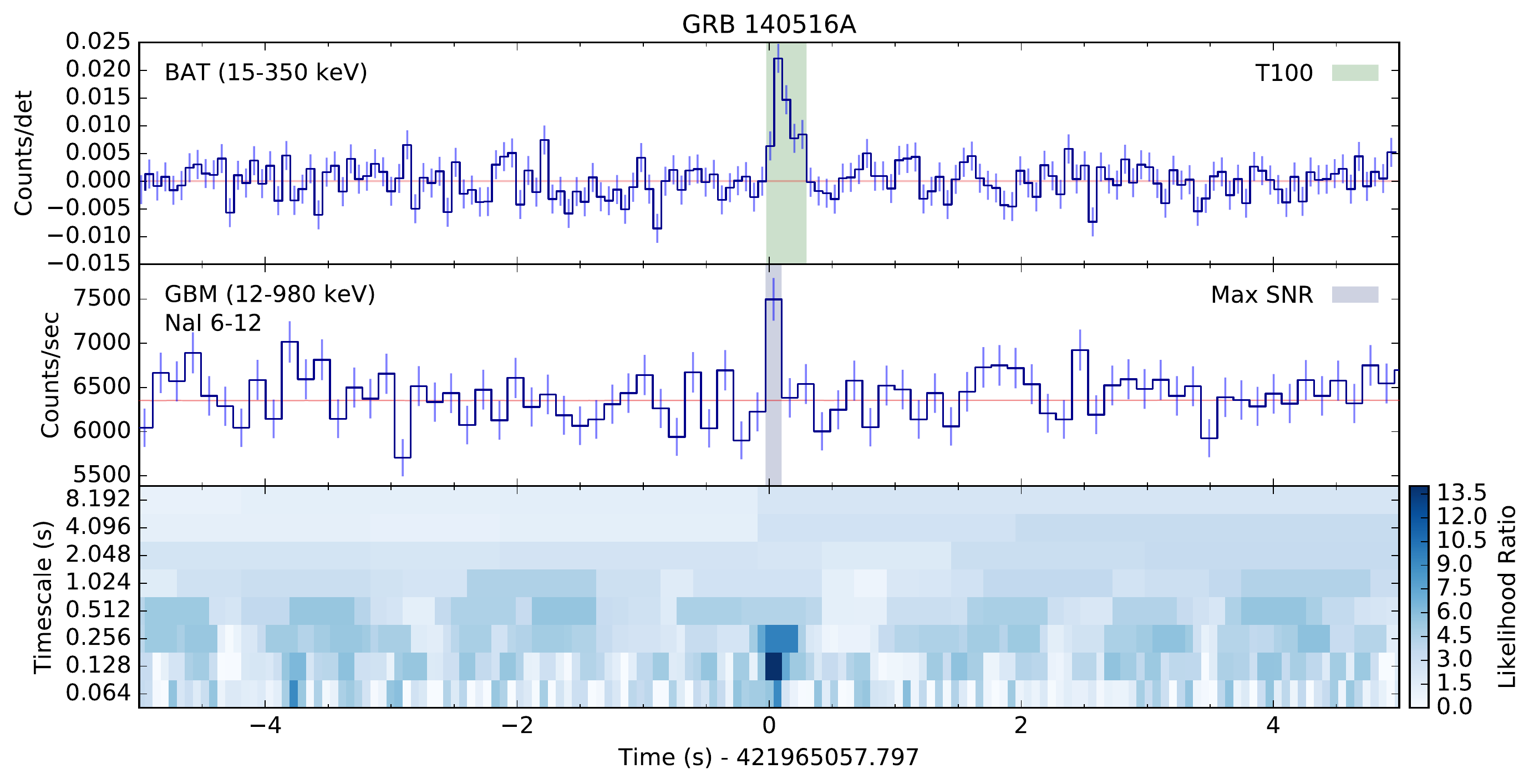}

\caption{Example BAT (top panels) and GBM (middle panels) light curves for 2 GRBs that only triggered BAT, but were recovered in the GBM CTTE data through ground analysis only on short timescales. The event display (lower panels) show the likelihood ratios, or the significance of the signal above the local background, for a range of bin timescales and phases.  The timescale that maximizes the signal significance in the GBM data is shown by the blue shaded region in the middle panels}
\label{Fig:LightCurves_UntriggeredExamples_ShortTimescales}
\end{figure}

\begin{figure}
\centering
\includegraphics[width=0.5\textwidth]{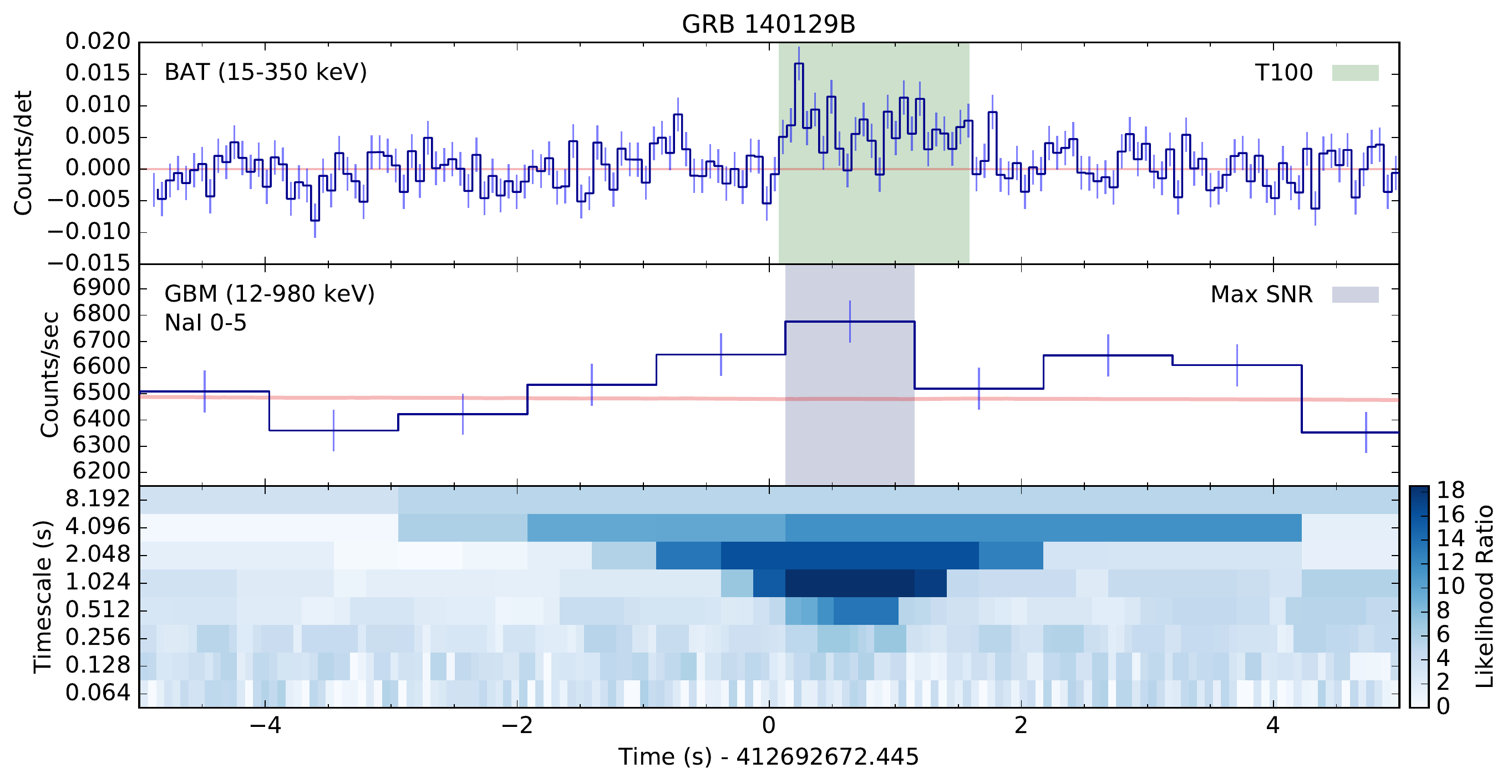}
\includegraphics[width=0.5\textwidth]{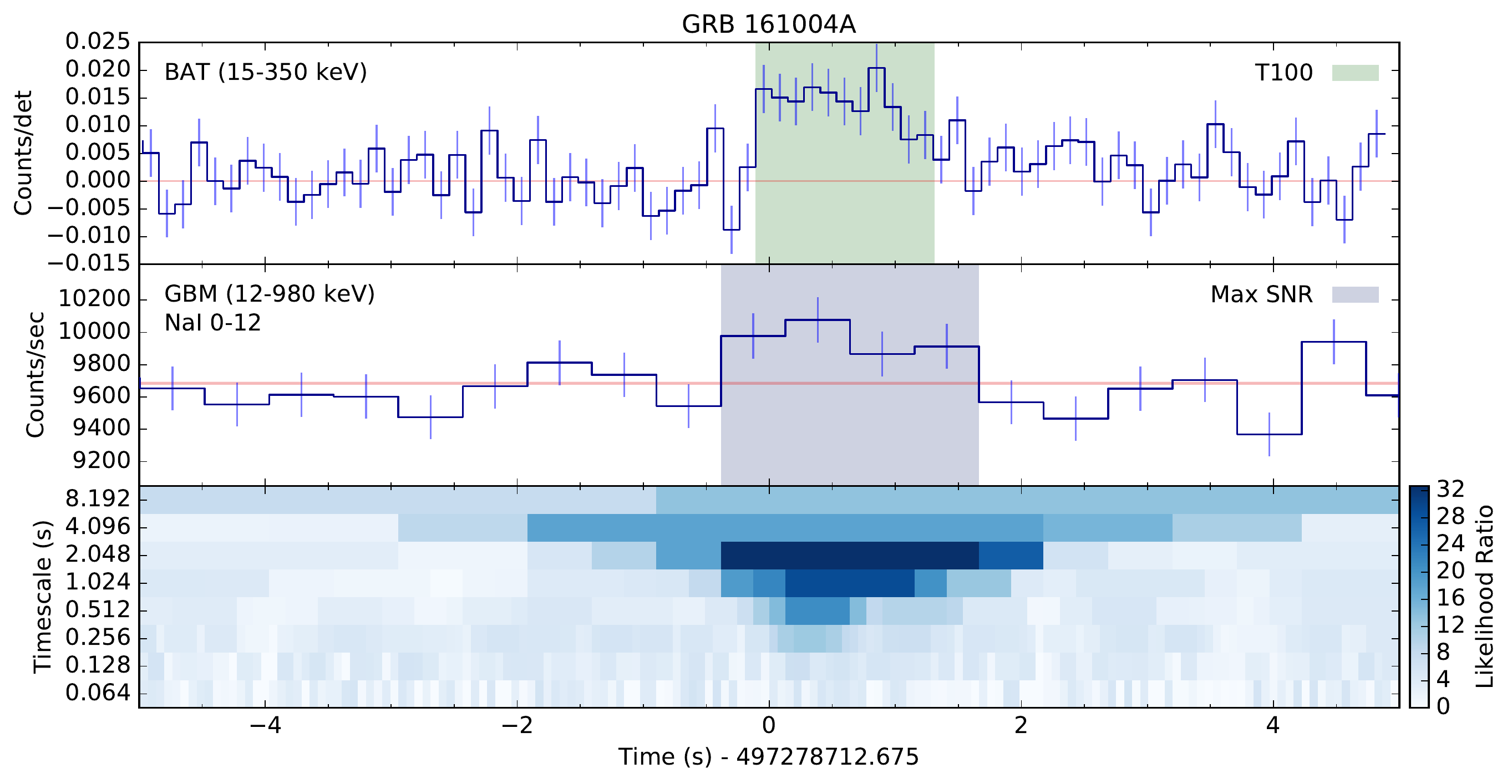}

\caption{Example BAT (top panels) and GBM (middle panels) light curves for 2 GRBs that only triggered BAT, but were recovered in the GBM CTTE data through ground analysis only on long timescales. The event display (lower panels) show the likelihood ratios, or the significance of the signal above the local background, for a range of bin timescales and phases.  The timescale that maximizes the signal significance in the GBM data is shown by the blue shaded region in the middle panels}
\label{Fig:LightCurves_UntriggeredExamples_LongTimescales}
\end{figure}

Similar plots for four bursts that only triggered BAT, but which were recoverable in the GBM CTTE data through ground analysis, are shown in Figures \ref{Fig:LightCurves_UntriggeredExamples_ShortTimescales} and \ref{Fig:LightCurves_UntriggeredExamples_LongTimescales}. Figure \ref{Fig:LightCurves_UntriggeredExamples_ShortTimescales} highlights two bursts that were recovered in the GBM data only on the shortest timescales analyzed by the search, whereas Figure \ref{Fig:LightCurves_UntriggeredExamples_LongTimescales} highlights two bursts that were detected only after integrating on longer timescales, reflecting the longer T$_{100}$ durations of the two bursts as seen by BAT.  All four bursts serve as examples of how the choice of binning can significantly affect the sensitivity to sub-threshold signals and highlights the need to examine the wide range of timescales and bin phasing that we employ. The lightcurve and event displays for GRB~170817A and GW150914-GBM are shown in Figure \ref{Fig:LightCurves_GWEvents} for comparison.

\begin{figure}
	\begin{center}
     \includegraphics[width=0.5\textwidth]{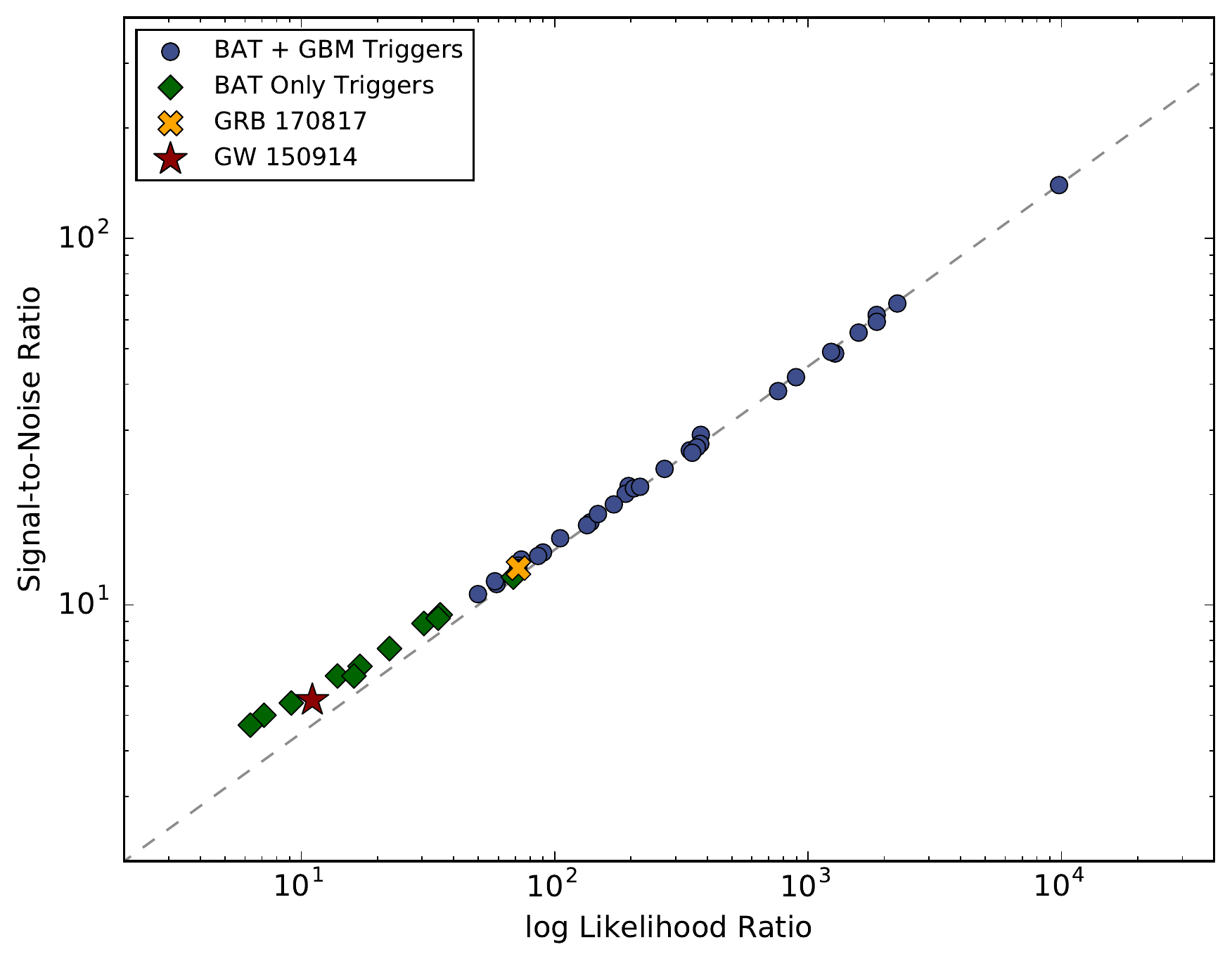}	
	\end{center}
\caption{The signal-to-noise of all 44 bursts in our sample versus their likelihood ratios. The circles represent the bursts that triggered both the BAT and GBM, while the diamonds are candidate GBM detections of bursts that only triggered BAT. The x and star represent GRB 170817A and GW150914-GBM. The gray dashed line represents the line of equivalence between the likelihood ratio and the square root of twice the SNR.}
\label{Fig:logLRvsSNR}
\end{figure}

The resulting SNR of all 44 bursts in our sample versus their likelihood ratios is shown in Figure \ref{Fig:logLRvsSNR}.  The gray dashed line represents the line of equivalence between the likelihood ratio and the square root of twice the SNR.  The bursts which triggered both \Swift BAT and \Fermi GBM (blue circles) and the GBM sub-threshold population (green diamonds) are well differentiated by both their likelihood ratio and SNR values.  Generally, bursts with likelihood ratios greater than $\mathcal{L} \gtrsim 40$ and SNR $\gtrsim 10$ resulted in on-board triggers of GBM, whereas the sub-threshold population are largely relegated to bursts with SNR$\lesssim 10$. 

\begin{figure}
	\begin{center}
     \includegraphics[width=0.5\textwidth]{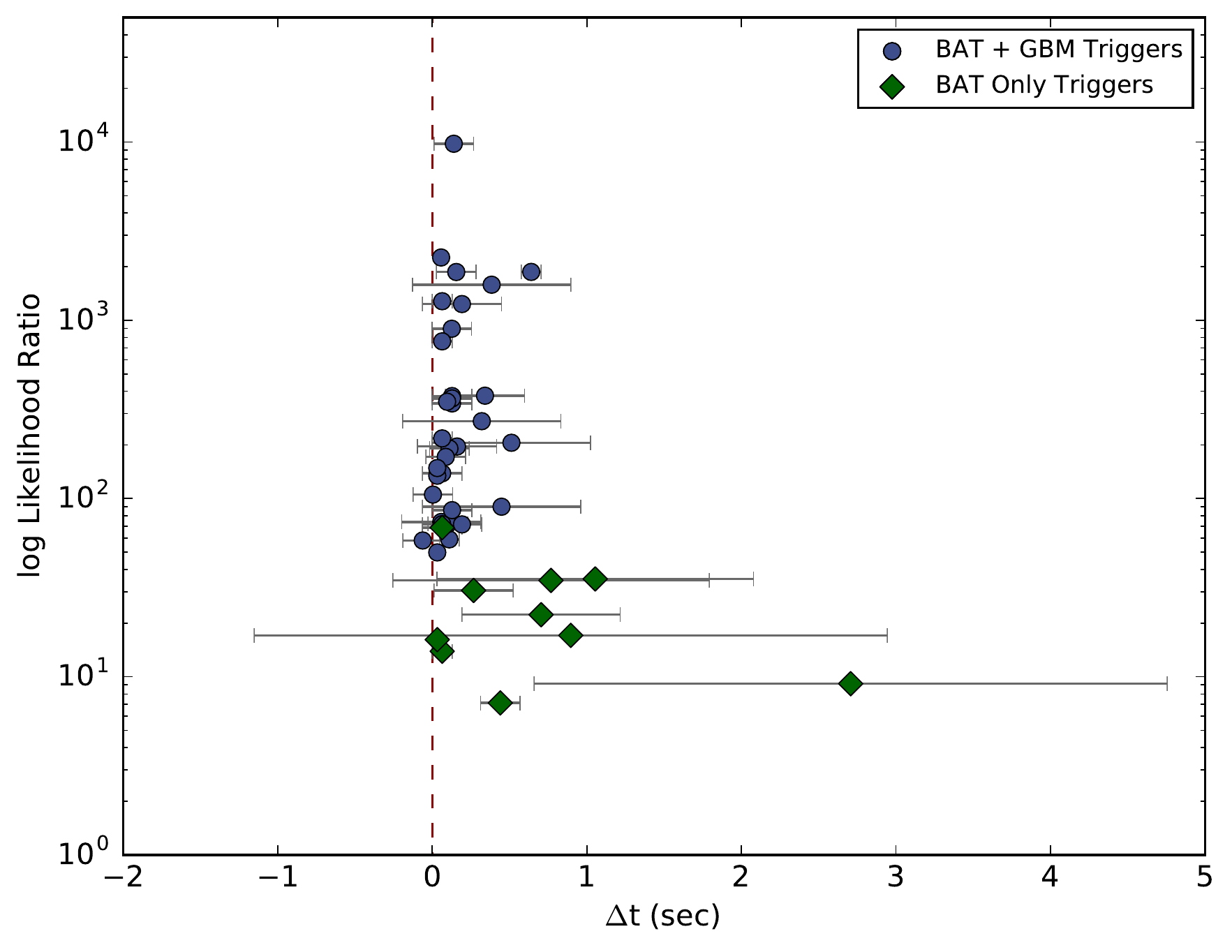}
	\end{center}
\caption{The likelihood ratio versus the offset between the BAT trigger time (T$_0$) and the center of the detection window that resulted in the highest signal significance for the bursts in our sample. The blue circles represent the bursts that triggered both the BAT and GBM and the green diamonds are candidate GBM detections of bursts that only triggered BAT.  The error bars represent the width of the detection window.}
\label{Fig:logLRvsTimeOffset}
\end{figure}

The likelihood ratio of the candidate detections versus their time offset $\Delta t$ from the BAT trigger time (T$_0$) is shown in Figure \ref{Fig:logLRvsTimeOffset}. Here the error bars represent the timescale of the source window on which the highest likelihood ratio was obtained by the targeted search.  The center of each timescale is systematically shifted to the right of the \Swift BAT T$_0$, as expected since the targeted search identifies the timescale that maximizes the significance of the signal over the background.  The BAT and GBM detected sample (blue circles) have a median offset of $\Delta t$ $\sim$ 0.126 s compared to the GBM sub-threshold population of $\Delta t$ $\sim$ 0.44 s.  

\begin{figure}
	\begin{center}
     \includegraphics[width=0.5\textwidth]{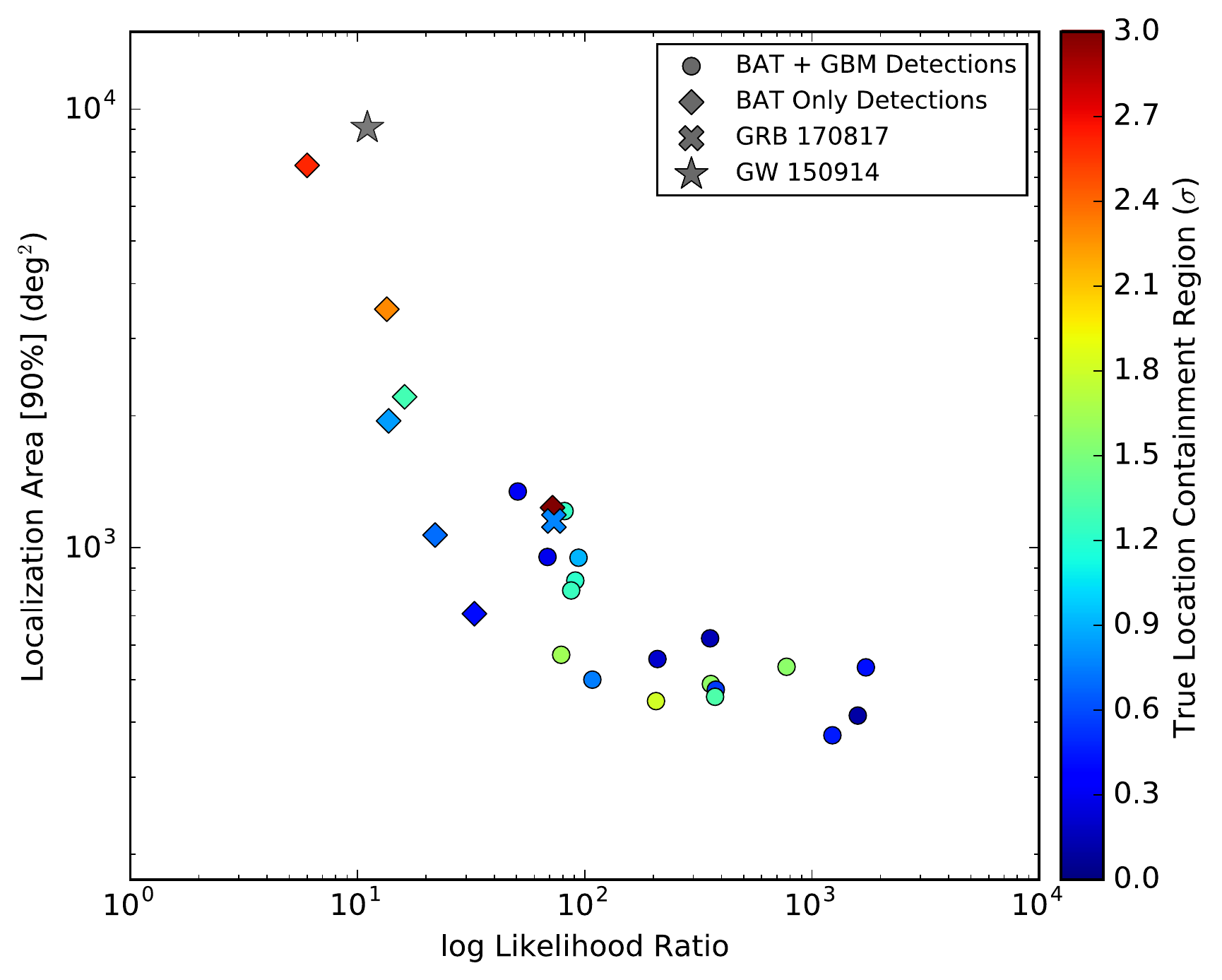}
	\end{center}
\caption{The GBM localization area (90$\%$ credible region) versus the likelihood ratio for each source in our sample with CTTE data.  The color scale represents the credible region of the GBM localization that is required to contain the true position of the source.  By definition, the position of GW150914-GBM is set to the maximum of the posterior probability distribution returned by the targeted search.}
\label{Fig:LocalizationVsLogLR}
\end{figure}

The area of the 90$\%$ localization credible region versus the likelihood ratio for each source is shown in Figure \ref{Fig:LocalizationVsLogLR}.  The color scale represents the credible region of the GBM localization that is required to contain the true position of the source.  The true position of GW150914-GBM is not known, so we exclude this event from the color scale, but plot the data point to compare its detection significance and localization area. Due to the present limitations of the targeted search when applied to legacy CTIME data, the localizations presented in Figure \ref{Fig:LocalizationVsLogLR} are limited to bursts after 2013 for which CTTE data is available. The resulting localization area is seen to fall sharply as a function of increasing detection significance, with a leveling off at high likelihood ratios due to the fixed 7.6$^\circ$ Gaussian systematic uncertainty added to each localization.  The true location of 84$\%$ of our sample, as determined by BAT, XRT, or optical detections, fall within the 90$\%$ ($\sim$1.65$\sigma$) credible region found by the targeted search.  The color scale reveals that the weakest bursts tend to disproportionately fall outside this region, indicating that the systematic uncertainty determined from triggered bursts may underestimate the true systematic error in the GBM localizations for the sub-threshold candidates.

The FAR versus the likelihood ratio for the entire SGRB sample is shown in Figure \ref{Fig:FAR}. The plot displays the rate at which transient signals of comparable significance to each burst are found while searching background intervals. Because of the different resolutions of the two data types, CTIME and CTTE, employed in this search, we constructed two separate FAR distributions using minimum timescales of 0.256 s and 0.064 s respectively.  Likewise, the FAR distribution is calculated separately for each of the three spectral templates employed by the search.  Overall, the rate at which background signals are detected, on both the 0.256 s and 0.064 s timescales and for the three different spectral templates, decreases as a function of their detection significance, as expected. The FAR analysis shows, though, that the rate at which soft transients are detected by chance in the GBM data far outweigh medium and hard signals of similar significance. Consequently, this decreases the probability of association since these types of signals occur by chance more often. Likewise, the plot also reveals that while running the search down to 0.064 s may help detect the shortest SGRBs in our sample, the rate of random occurrence at this timescales is substantial.  For example, a search of CTTE data at the 0.064 s timescale can be expected to detect a soft transient with $\mathcal{L} \sim 10$ once every 1/3.5$\times 10^{-3}$ $\sim$ 211 s.  This dramatically decreases the post-trials significance of signals detected using the soft template on the shortest timescale employed by the search.

Finally, the likelihood ratio of the SGRB sample versus their energy fluence, in the 15$-$350 keV energy range as measured by BAT, is shown in Figure \ref{Fig:Fluence_15_350_Best_T100}.  Here the energy fluence is measured using the best fit spectral model integrated over the observed T$_{100}$ duration, which is represented as the color of each data point.  The likelihood ratio roughly correlates with the burst fluence, with bursts of similar fluence but shorter durations, and hence higher peak flux, generally result in higher likelihood ratios.  In addition to the Swift detected SGRB sample, Figure \ref{Fig:Fluence_15_350_Best_T100} also includes the likelihood ratio and energy fluence, in the equivalent 15$-$350 keV energy range as measured by GBM, for GRB 170817A and GW150914-GBM. 

The results of the targeted search for the entire sample, ordered by decreasing likelihood ratio, are displayed in Table \ref{table:Sample}.

\section{Discussion}

The results outlined in Section 5 reveal that the GBM targeted search can be an effective method of identifying weak transient signals hidden in the untriggered GBM CTTE data. By taking into account the viewing geometry and response of each detector, the targeted search capitalizes on the increased sensitivity that is obtained through a coherent stacking of the multi-detector data.  We find that all but two bursts in our sample resulted in a candidate signal above our pre-trial significance of $3\sigma$ which was also consistent with the BAT trigger time.

Figure 4 reveals that a signal to noise threshold of roughly SNR$\sim$10 separates the BAT detected bursts that also triggered GBM and those that were only recovered through the targeted search. One notable exception to this delineation is GRB 140606A, which was well detected by the targeted search, with a likelihood ratio of $\mathcal{L}\sim69$ and a SNR~$\sim$12. This burst appears to have occurred in a unique position in spacecraft coordinates in which it brightly illuminated only one NaI detector, thus failing the on-board triggering criteria of a 4.5$\sigma$ rate increase in at least two detectors. Since a statistically significant signal in at least two detectors is not a criteria for detection by the targeted search, these events are easily recovered through ground analysis. 

The SNR and likelihood ratio for GRB 170817A is also included in Figure \ref{Fig:logLRvsSNR}, showing the proximity of the burst to the GBM on-board detection threshold.  Detected with SNR$\sim$12.7, \citet{Goldstein2017} estimated that GRB~170817A could have been at most $\sim30\%$ dimmer and still have triggered GBM, assuming the same background and viewing geometry.  This corresponds to roughly SNR$\sim$9, consistent with the empirical determination of the on-board detection threshold found from the Swift detected SGRB sample.  According to this analysis, the targeted search could have detected GRB 170817A with $\mathcal{L}>9$ and SNR$>5$ if the burst had been dimmed by roughly $60\%$ of its original brightness, consistent to the conclusions drawn by \citet{GBM_only_paper}. Having been detected at a distance of 43 Mpc with a flux of $\mathcal{F} = 3.7 \pm 0.9$ ph s$^{-1}$ cm$^{-2}$ (50--300 keV), this corresponds to a $\sim 60\%$ increase in the maximum detection distance to $\sim$ 74 Mpc. Therefore the increased sensitivity achieved through the targeted search will be crucial to expanding the gamma-ray horizon at which such events can be detected by GBM. While this increase does not seem overwhelming, the GW detection rate goes as the volume searched, corresponding to an increase of a factor of 5 in the volume of the Universe in which GRB~170817A could have been detected by the targeted search. 

\begin{figure}
	\begin{center}
     \includegraphics[width=0.5\textwidth]{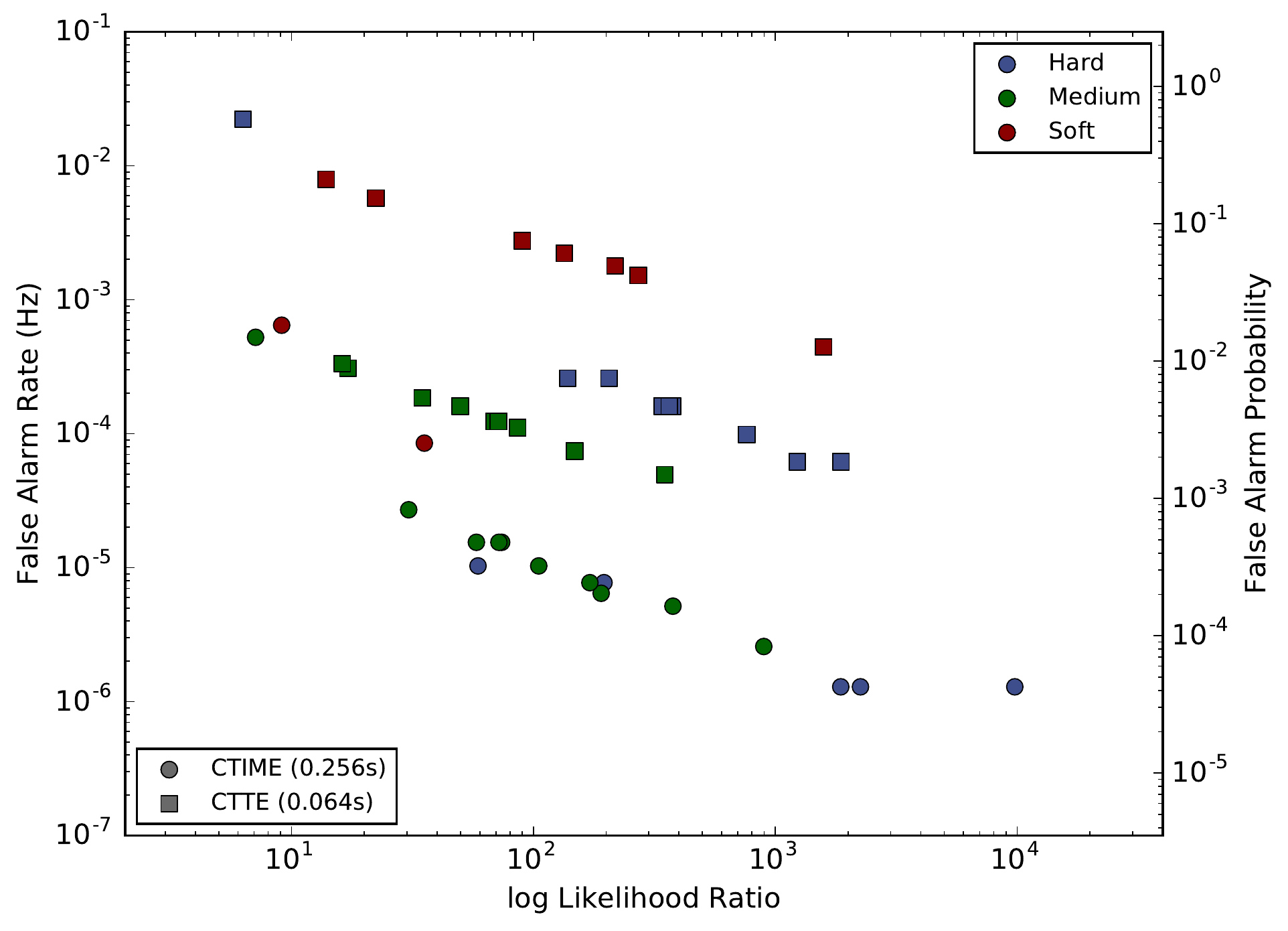}
	\end{center}
\caption{The FAR versus the likelihood ratio for the entire SGRB sample displaying the rate at which transient signals of comparable significance to each burst are found while searching random intervals of GBM data.}
\label{Fig:FAR}
\end{figure}

The FAR results presented in Figure \ref{Fig:FAR} reveal that while pushing the resolution of targeted search down to 0.064 s may aid in the detection of particularly short signals, the probability of chance coincidence with background transients (of either statistical or astrophysical nature) becomes substantial. This is especially true of source detected using the soft spectral template, likely due to a background of soft transients of galactic origin being picked up in the data.  Therefore, the increased sensitivity to the shortest signals when running the search at such fine timescales comes at the cost of association significance. The conclusions drawn from the analysis of this control sample have important implications for the optimization of the targeted search for the search of EM counterparts to GW detections. 

\begin{figure}
	\begin{center}
     \includegraphics[width=0.5\textwidth]{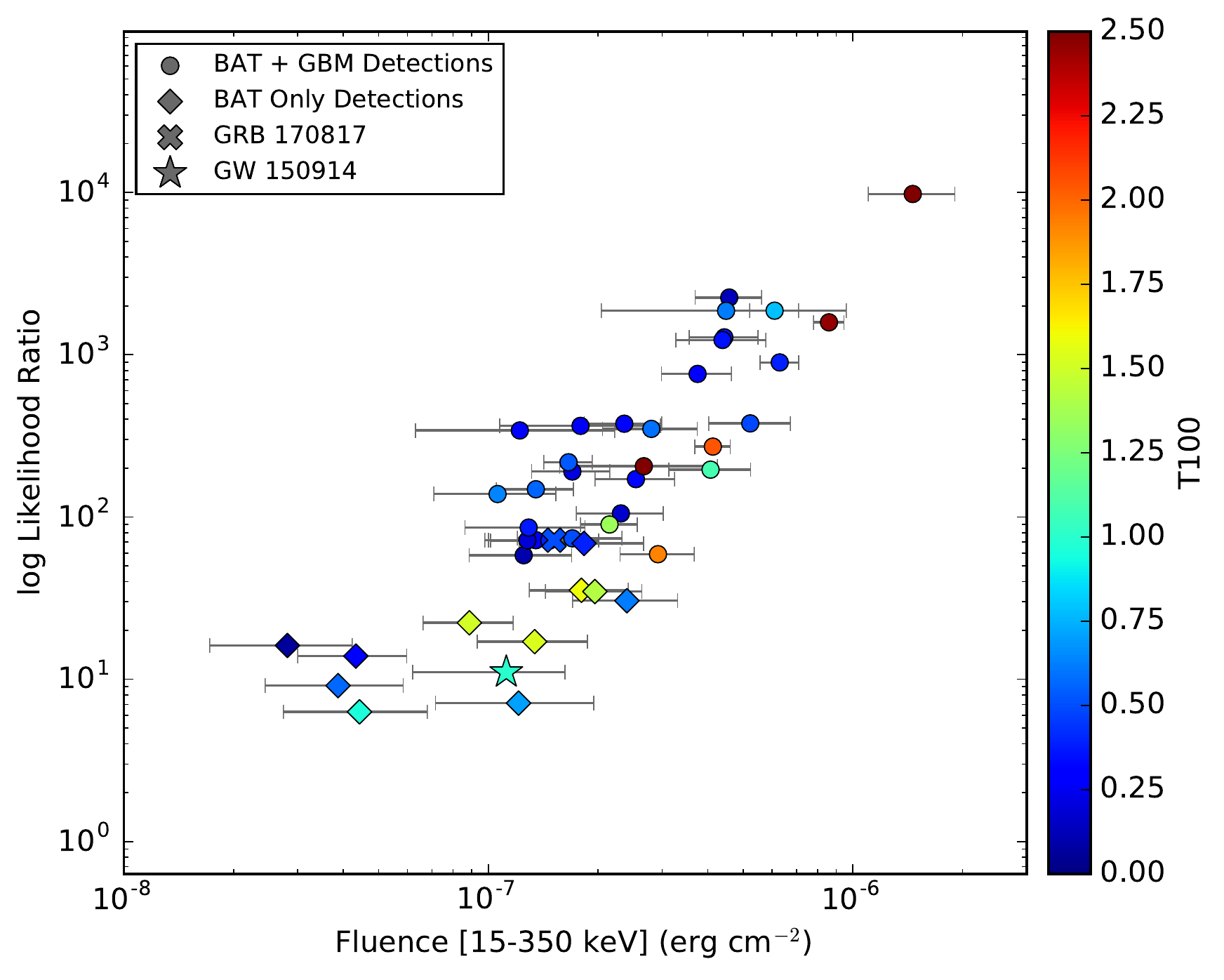}	
	\end{center}
\caption{The likelihood ratio versus fluence as measured by BAT in the 15-350 keV energy range for the bursts in our sample. The color of each data point represents the BAT measured T$_{100}$ duration.  The circles represent the bursts that triggered both the BAT and GBM, while the diamonds are candidate GBM detections of bursts that only triggered BAT.  The x and star represent the fluence in the same energy range as measured by GBM for GRB 170817A and GW150914-GBM.}
\label{Fig:Fluence_15_350_Best_T100}
\end{figure}

The likelihood ratio of the SGRB sample versus their energy fluence shown in Figure \ref{Fig:Fluence_15_350_Best_T100} reveals that the triggered and sub-threshold populations are not easily delineated by a single physical fluence threshold.  Instead, burst fluence, duration, and viewing geometry all play a role in the chance that a burst would trigger the GBM. A general trend is evident in Figure \ref{Fig:Fluence_15_350_Best_T100} in which bursts of similar fluence values, but of shorter duration, tend to yield higher likelihood ratios and are generally more likely to have resulted in an on-board trigger of the instrument.  This can be understood in the context of the on-board rate trigger responding to the peak flux of the burst, rather than the total fluence spread over the duration of the burst. At the same time, GRB~140606A did not result in an on-board trigger, whereas GRB~170817A did, despite being of comparable fluence and duration and yielding identical likelihood ratio values. The difference between the two bursts was that GRB~140606A occurred at a location in spacecraft coordinates that precluded the bright illumination of all but one of the NaI detectors. The two bursts highlight the importance of viewing geometry, in addition to intrinsic flux and fluence, in the on-board detectability of a burst.

\begin{figure}
\centering
\includegraphics[width=0.5\textwidth]{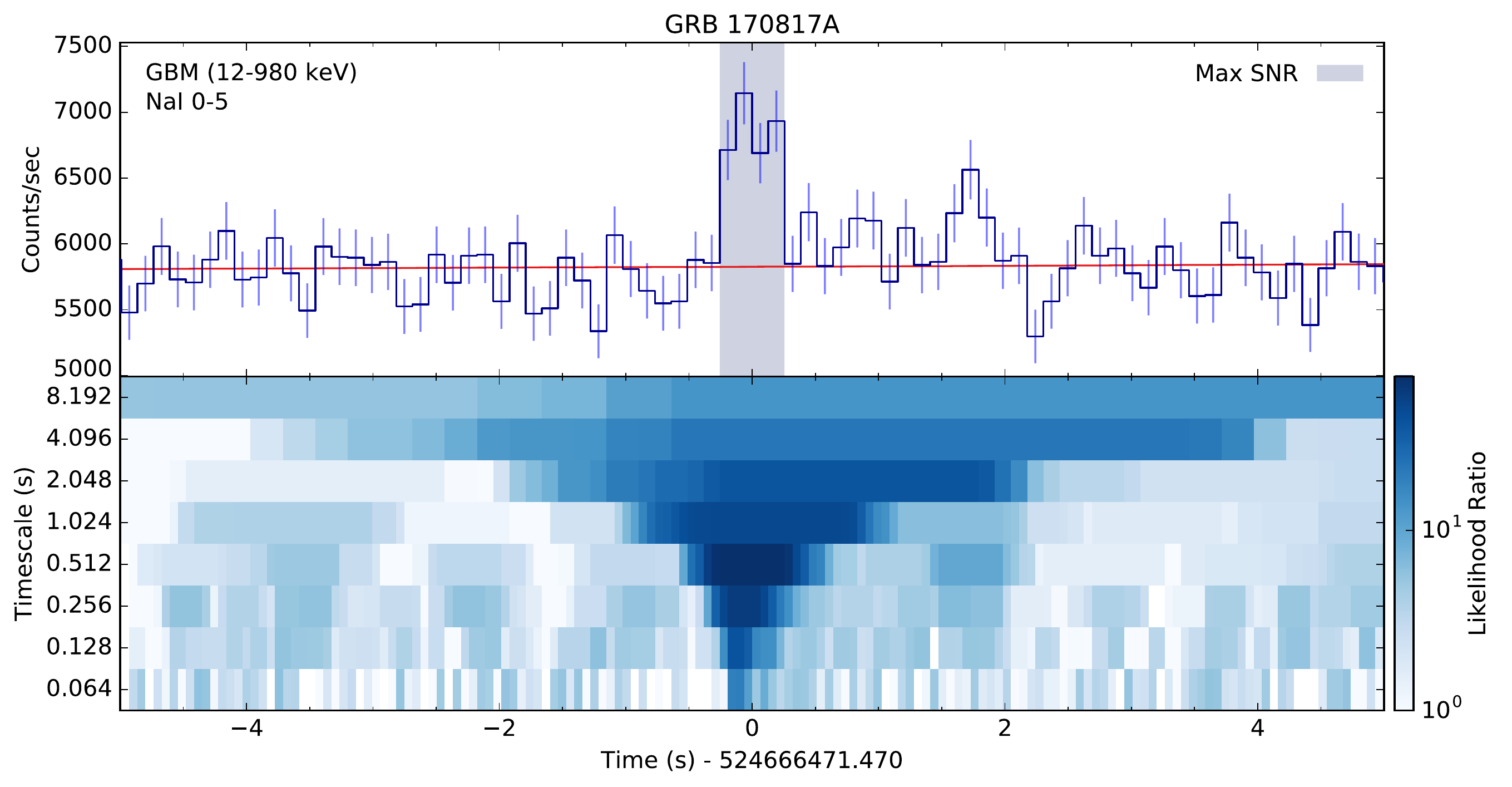}
\includegraphics[width=0.5\textwidth]{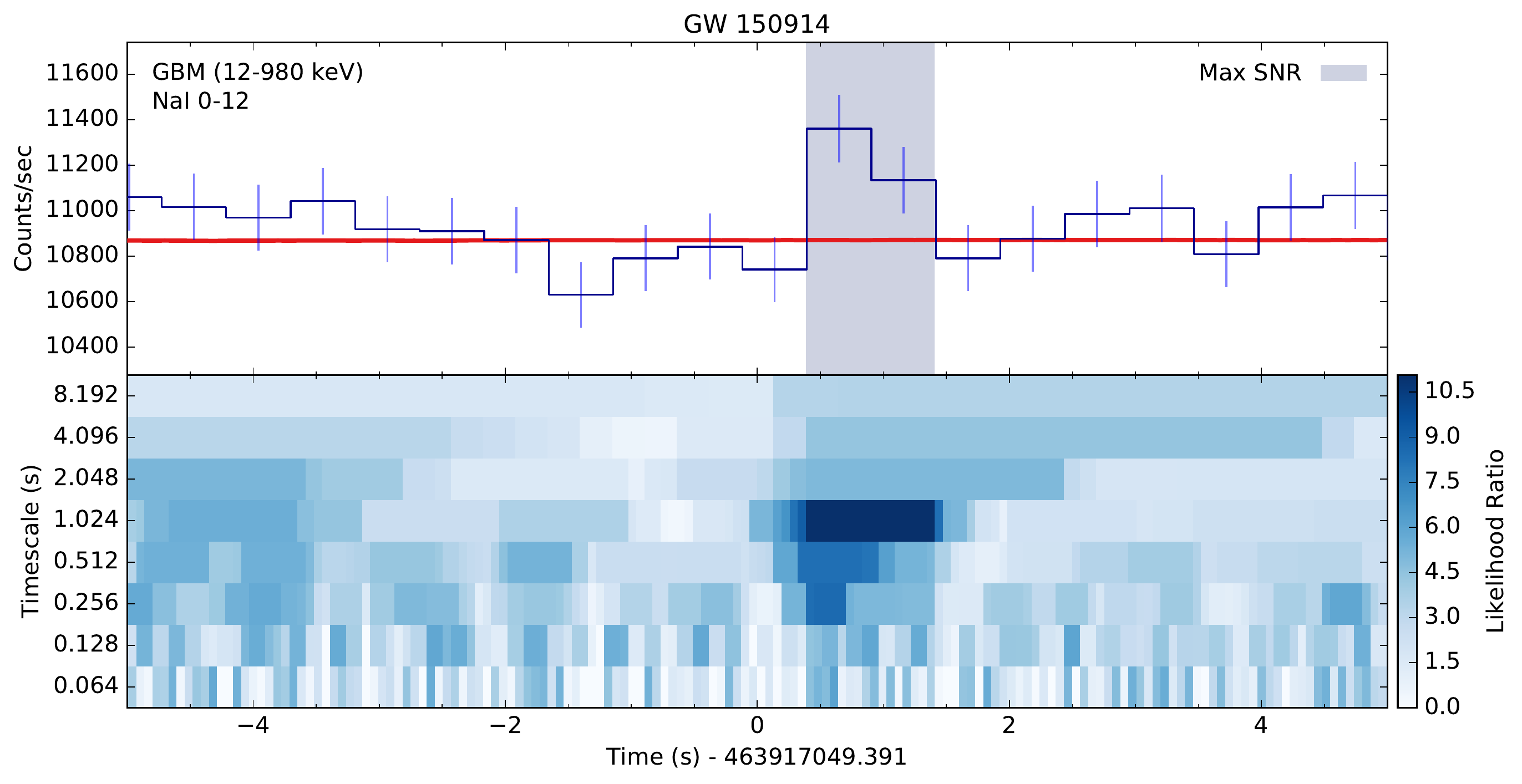}
\caption{The GBM light curves for GRB~170817A and GW150914-GBM. The event display (lower panels) show the likelihood ratios, or the significance of the signal above the local background, for a range of bin timescales and phases.  The timescale that maximizes the signal significance in the GBM data is shown by the blue shaded region in the top panels.}
\label{Fig:LightCurves_GWEvents}
\end{figure}

The location of GW150914-GBM in Figure \ref{Fig:Fluence_15_350_Best_T100} is also revealing. The GW150914-GBM candidate was identified 0.4 s after the LIGO trigger, with a likelihood ratio of $\sim11$ and SNR$\sim5.5$.  This places GW150914-GBM near the detection threshold we have adopted for the targeted search, but squarely within the GBM sub-threshold population of astrophysical SGRBs detected by BAT. The comparison shows that the targeted search is capable of recovering astrophysical signals as weak as GW150914-GBM in the untriggered data. The GBM light curve and event analysis display for GW150914-GBM are shown in Figure \ref{Fig:LightCurves_GWEvents}, revealing a weak $\sim$ 1 s long source that was detected on multiple timescales. 

The nature of GW150914-GBM has resulted in a vigorous debate in the gamma-ray and GW communities.  The original investigation of the GBM data by \citet{Connaughton2016} revealed a weak source with a location consistent with the LIGO localization of GW150914 and a hard spectrum typical of SGRBs. Based on the proximity in time of the signal to the GW event and the rate of transients of similar significance in the GBM data, \citet{Connaughton2016} estimated a post-trials association probability of 2.9$\sigma$.  The source was not detected by INTEGRAL SPI-ACS \citealt{savchenko2016integral}, which viewed the GBM localization, and the statistical significance of the signal was challenged by \citealt{greiner2016fermi}.  We note that INTEGRAL SPI-ACS is sensitive to higher energies than the GBM and the large uncertainty in the GBM inferred spectrum renders the INTEGRAL non-detection inconclusive. \citealt{2018arXiv180102305C} has also challenged the conclusions drawn by \citealt{greiner2016fermi} and pointed out claimed misrepresentations made by \citealt{greiner2016fermi} of the analysis originally performed in \citet{Connaughton2016}.

The low association probability notwithstanding, perhaps the greatest challenge facing GW150914-GBM has been that merging black holes in vacuum are not expected to generate electromagnetic signals.  This has not prevented a series of authors from devising a list of possible scenarios to explain the origin of the signal.  These range from the dissipation of the Poynting flux energy through the merger of two charged black holes \citep{Zhang2016, Fraschetti2016, Liebling2016}; super-Eddington accretion through a BBH merger within an AGN disk \citep{Bartos2017, Stone2017}, or a BBH system with a long-lived disk of ambient material \citep{Perna2016, Murase2016}.  Other theories include the triggered collapse of a massive star due to its black hole companion \citep{Janiuk2017}, and the fragmentation of the stellar core of a massive star \citep{Loeb2016} (although see \citet{Dai2017} and \citet{Fedrow2017}).  \citet{Lyutikov2016} has argued that the physical properties necessary to create GW150914-GBM are highly implausible, although \citet{Khan2018} surveyed different models of disks around merging BBHs using magnetohydrodynamic simulations and shows that such systems could produce an EM counterpart consistent with the properties of GW150914-GBM.  Likewise \citet{Veres2016} showed that dissipative photosphere models of GRB emission can accommodate the GBM observations.

Ultimately, the detection of another EM signal from a BBH system will be required to  conclusively settle the debate as to the origin of GW150914-GBM.  We show that the offline targeted search developed by the GBM team is capable of recovering true astrophysical signals as weak and/or viewed with poor spacecraft geometry as GW150914-GBM in the untriggered data. The use of this technique to examine GBM observations of future BBH merger candidates detected by LIGO/Virgo will be crucial at resolving this mystery.

\section{Conclusions}

We have performed an extensive examination of the efficiency with which the GBM offline targeted search can recover true astrophysical signals in the GBM CTTE data.  This method was originally developed by \citet{Blackburn2015} and extended by \citet{Goldstein2016} to search CTTE data for sub-threshold transient events in temporal coincidence with a LIGO/Virgo compact binary coalescence triggers and plays a crucial role in the GBM follow-up of GW events. This targeted search operates by coherently combining data from all 14 GBM detectors by taking into account the complex spatial and energy dependent response of each detector. We use the method to examine a sample of SGRBs that were independently detected by the \Swift BAT, but which were too intrinsically weak, distant, or viewed with poor spacecraft geometry to initiate an on-board trigger of the GBM.  We find that the search can successfully recover a majority of the BAT detected sample in the CTTE data. We show that the targeted search of CTTE data is crucial to increasing the GBM sensitivity, and hence the gamma-ray horizon, to events such as GRB~170817A.  We find that the increased sensitivity of the targeted search results in a $\sim 60\%$ increase in the maximum detection distance of GRB~170817A. Since the detection rate scales as the volume searched, this corresponds to a 5 fold increase in the volume of the Universe in which GRB~170817A could have been detected by the targeted search.  Finally, we also examine the properties of the GBM signal possibly associated with the LIGO detection of GW150914.  We show that the signal is consistent with the observed properties of the GBM sub-threshold population of astrophysical SGRBs detected by BAT that are recoverable through the offline targeted search. Despite new theories that could explain the origin of the GW150914-GBM, we conclude that future detections of EM signals from BBH merger candidates detected by LIGO/Virgo will be needed to confirm the astrophysical nature of GW150914-GBM.

\bibliographystyle{aasjournal}
\bibliography{bibliography.bib}

\newpage
\newpage

\LongTables
\begin{deluxetable*}{ccccccccccccc}
\tablecaption{Sample Definition $\&$ Summary of Results}
\tablehead{
\colhead{GRB} & \colhead{MET\tablenotemark{~a,}\tablenotemark{~b}} & \colhead{$\Delta$t} & \colhead{Timescale} & \colhead{$\mathcal{L}$}  & \colhead{Fluence\tablenotemark{~c}} & \colhead{FAR} & \colhead{FAP} & \colhead{$\theta$\tablenotemark{~d}} & \colhead{$\phi$\tablenotemark{~e}} & \colhead{$\psi_{1}$\tablenotemark{f}} & \colhead{$\psi_{2}$\tablenotemark{g}} & \colhead{Template}\\ [0.5ex] 
 & \colhead{(s)} & \colhead{(s)} & \colhead{(s)} & \colhead{} & \colhead{(erg cm$^{-2}$)} & \colhead{(Hz)} & \colhead{$p$} & \colhead{(deg)} & \colhead{(deg)} & \colhead{(deg)} & \colhead{(deg)} & \colhead{}
}
\startdata
090510A & 263607782.488 & 0.14 & 0.256 & 9780.1 & 1.46e-06 & 1.29e-06 & 4.00e-05 & 13.6 & 231.6 & 7.1 & 32.7 & medium \\
100206A & 287155807.374 & 0.06 & 0.064 & 2250.4 & 4.58e-07 & 1.29e-06 & 4.00e-05 & 44.6 & 14.1 & 21.8 & 28.8 & medium \\
160726A & 491189651.673 & 0.64 & 0.128 & 1867.8 & 6.10e-07 & 2.97e-05 & 8.90e-04 & 44.2 & 53.2 & 5.8 & 23.9 & soft \\
101129A & 312737973.648 & 0.15 & 0.256 & 1865.3 & 4.49e-07 & 1.29e-06 & 4.00e-05 & 25.8 & 114.1 & 23.0 & 25.9 & soft \\
151229A & 473064631.963 & 0.38 & 1.024 & 1583.0 & 8.60e-07 & 1.19e-04 & 3.56e-03 & 53.4 & 59.6 & 13.6 & 33.7 & medium \\
130515A & 390273680.84 & 0.06 & 0.128 & 1279.1 & 4.44e-07 & 2.97e-05 & 8.90e-04 & 128.3 & 260.7 & 44.3 & 54.4 & hard \\
160408A & 481789547.777 & 0.19 & 0.512 & 1233.0 & 4.39e-07 & 2.97e-05 & 8.90e-04 & 6.7 & 53.8 & 14.0 & 27.1 & medium \\
100625A & 299183550.378 & 0.13 & 0.256 & 896.3 & 6.30e-07 & 2.58e-06 & 8.00e-05 & 124.2 & 291.2 & 35.8 & 61.4 & soft \\
130912A & 400667700.933 & 0.06 & 0.128 & 761.6 & 3.75e-07 & 2.97e-05 & 8.90e-04 & 102.0 & 229.5 & 13.9 & 46.9 & hard \\
111117A & 343224823.922 & 0.34 & 0.512 & 377.4 & 5.23e-07 & 5.16e-06 & 1.50e-04 & 13.6 & 114.2 & 19.9 & 28.2 & hard \\
151228A & 472964716.385 & 0.13 & 0.256 & 375.6 & 2.36e-07 & 7.43e-05 & 2.23e-03 & 117.1 & 288.9 & 30.1 & 57.0 & medium \\
160624A & 488460425.298 & 0.13 & 0.256 & 364.3 & 1.79e-07 & 7.43e-05 & 2.23e-03 & 75.8 & 94.3 & 32.6 & 38.3 & medium \\
170127B & 507222813.676 & 0.10 & 0.064 & 349.1 & 2.80e-07 & 2.97e-05 & 8.90e-04 & 44.1 & 304.7 & 7.2 & 44.3 & medium \\
140402A & 418090209.854 & 0.13 & 0.256 & 341.5 & 1.22e-07 & 7.43e-05 & 2.23e-03 & 13.6 & 295.5 & 20.3 & 28.2 & hard \\
131004A & 402615666.688 & 0.32 & 1.024 & 271.5 & 4.13e-07 & 5.50e-04 & 1.64e-02 & 93.1 & 116.0 & 8.2 & 50.6 & medium \\
160821B & 493511357.303 & 0.06 & 0.128 & 217.4 & 1.66e-07 & 7.14e-04 & 2.12e-02 & 60.8 & 171.6 & 31.7 & 32.5 & soft \\
141205A & 439459520.172 & 0.51 & 1.024 & 205.5 & 2.67e-07 & 1.64e-04 & 4.89e-03 & 122.7 & 165.5 & 36.7 & 50.9 & medium \\
080905A & 242308735.974 & 0.16 & 0.512 & 195.8 & 4.07e-07 & 7.73e-06 & 2.30e-04 & 27.9 & 262.9 & 16.9 & 28.7 & medium \\
081101A & 247232792.979 & 0.11 & 0.256 & 190.6 & 1.70e-07 & 6.44e-06 & 1.90e-04 & 29.5 & 148.1 & 17.8 & 31.3 & hard \\
100117A & 285455181.576 & 0.09 & 0.256 & 171.2 & 2.54e-07 & 7.73e-06 & 2.30e-04 & 86.0 & 289.4 & 14.4 & 46.4 & hard \\
150301A & 446864671.635 & 0.03 & 0.064 & 148.2 & 1.35e-07 & 5.95e-05 & 1.78e-03 & 105.5 & 276.4 & 30.5 & 42.3 & medium \\
140320A & 416974368.947 & 0.06 & 0.256 & 138.5 & 1.06e-07 & 1.64e-04 & 4.89e-03 & 28.4 & 281.7 & 23.7 & 25.6 & medium \\
150101B & 441818617.459 & 0.03 & 0.064 & 134.2 & 1.40e-09 & 9.66e-04 & 2.86e-02 & 54.9 & 260.5 & 28.8 & 39.5 & hard \\
090621B & 267314847.649 & 0.00 & 0.256 & 105.2 & 2.31e-07 & 1.03e-05 & 3.10e-04 & 110.4 & 50.6 & 21.6 & 50.6 & hard \\
150120A & 443415469.413 & 0.45 & 1.024 & 90.0 & 2.15e-07 & 1.23e-03 & 3.63e-02 & 98.5 & 340.0 & 24.9 & 37.6 & medium \\
160714A & 490155559.539 & 0.13 & 0.256 & 86.0 & 1.29e-07 & 7.43e-05 & 2.23e-03 & 58.4 & 167.0 & 27.9 & 35.6 & soft \\
081226A & 251946218.372 & 0.06 & 0.512 & 73.9 & 1.70e-07 & 1.55e-05 & 4.60e-04 & 113.4 & 79.7 & 31.0 & 48.5 & medium \\
101224A & 314861235.761 & 0.10 & 0.256 & 72.0 & 1.35e-07 & 1.55e-05 & 4.60e-04 & 72.9 & 346.8 & 23.5 & 38.4 & medium \\
130626A & 393936666.777 & 0.06 & 0.256 & 71.8 & 1.28e-07 & 7.43e-05 & 2.23e-03 & 88.7 & 180.1 & 4.0 & 56.4 & medium \\
160411A & 482030935.924 & 0.19 & 0.256 & 71.6 & 1.47e-07 & 7.43e-05 & 2.23e-03 & 107.8 & 303.2 & 17.6 & 61.7 & hard \\
140606A$\dagger$ & 423745096.496 & 0.06 & 0.256 & 68.7 & 1.83e-07 & 7.43e-05 & 2.23e-03 & 91.9 & 254.2 & 17.7 & 49.0 & hard \\
081024A & 246520389.865 & 0.11 & 0.128 & 59.0 & 2.92e-07 & 1.03e-05 & 3.10e-04 & 120.6 & 184.0 & 30.3 & 58.5 & soft \\
110420B & 325032133.716 & -0.06 & 0.256 & 58.1 & 1.25e-07 & 1.55e-05 & 4.60e-04 & 123.2 & 16.7 & 35.7 & 51.2 & hard \\
150101A & 441786536.699 & 0.03 & 0.064 & 49.8 & 3.23e-09 & 1.04e-04 & 3.12e-03 & 12.4 & 348.6 & 17.2 & 29.1 & medium \\
120403A$\dagger$ & 355107925.588 & 1.05 & 2.048 & 35.4 & 1.80e-07 & 8.51e-05 & 2.55e-03 & 71.6 & 335.6 & 31.5 & 32.7 & soft \\
161004A$\dagger$ & 497278712.675 & 0.77 & 2.048 & 34.7 & 1.96e-07 & 1.19e-04 & 3.56e-03 & 158.5 & 98.3 & 70.3 & 73.4 & soft \\
090305A$\dagger$ & 257923193.035 & 0.27 & 0.512 & 30.5 & 2.40e-07 & 2.71e-05 & 8.10e-04 & 97.1 & 264.5 & 28.7 & 39.2 & hard \\
140129B$\dagger$ & 412692672.445 & 0.70 & 1.024 & 22.3 & 8.87e-08 & 3.09e-03 & 8.86e-02 & 13.9 & 17.3 & 10.6 & 33.3 & medium \\
151205B$\dagger$ & 471044598.536 & 0.90 & 4.096 & 17.1 & 1.34e-07 & 1.93e-04 & 5.78e-03 & 103.4 & 241.6 & 14.3 & 58.7 & hard \\
170112A$\dagger$ & 505879325.113 & 0.03 & 0.064 & 16.1 & 2.81e-08 & 2.08e-04 & 6.22e-03 & 89.9 & 74.7 & 16.2 & 49.1 & medium \\
140516A$\dagger$ & 421965057.797 & 0.06 & 0.128 & 13.9 & 4.33e-08 & 4.55e-03 & 1.28e-01 & 31.5 & 153.1 & 17.8 & 31.1 & medium \\
110112A$\dagger$ & 316498340.981 & 2.71 & 4.096 & 9.1 & 3.87e-08 & 6.46e-04 & 1.92e-02 & 134.6 & 292.4 & 45.3 & 66.4 & hard \\
090815C$\dagger$ & 272071301.223 & 0.44 & 0.256 & 7.1 & 1.21e-07 & 5.25e-04 & 1.56e-02 & 116.2 & 38.6 & 32.3 & 43.0 & soft \\
150728A$\dagger$ & 459780675.037 & -2.78 & 0.064 & 6.3 & 4.43e-08 & 2.07e-02 & 4.62e-01 & 19.6 & 79.1 & 11.3 & 30.7 & hard \\
\enddata
\tablenotetext{a}{The \Fermi mission elapsed time (MET) representing the number of seconds since January 1, 2001, at 00:00:00 UTC.}
\tablenotetext{a}{The \Fermi MET corresponding to the \Swift BAT trigger time.}
\tablenotetext{c}{Burst fluence values reproduced from \citet{Lien2016}}
\tablenotetext{d}{The zenith angle from the spacecraft boresight to the burst position.}
\tablenotetext{e}{The azimuth angle from the spacecraft +X axis (sun facing) to the burst position.}
\tablenotetext{f}{The angle between the pointing direction of the closest GBM detector to the burst position.}
\tablenotetext{g}{The angle between the pointing direction of the second closest GBM detector to the burst position.}
\tablecomments{The results for the 44 BAT detected SGRBs in our sample, ordered by decreasing $\mathcal{L}$.  Bursts denoted by a $\dagger$ represent the events which were detected by BAT, but did not initiate an onboard trigger of GBM.}
\label{table:Sample}
\end{deluxetable*}

\end{document}